\documentclass[twocolumn,showpacs,aps,prl,groupedaddress]{revtex4}
\usepackage{epsfig,amssymb,amsmath}

\def\labell#1{\label{#1}}
\def\section#1{{\em #1:--- }}

\def\togli#1{}
\newcommand {\rhovec}{\ensuremath \boldsymbol{\rho}}
\newcommand {\thetavec}{\ensuremath \boldsymbol{\theta}}
\begin{document}
\title{Sub-Rayleigh Imaging via $N$-Photon Detection}

\author{Fabrizio Guerrieri$^{1,2}$, Lorenzo Maccone$^2$, Franco N. C.
  Wong$^2$, Jeffrey H. Shapiro$^2$, Simone Tisa$^3$, and Franco
  Zappa$^1$}
\affiliation{$^1$Politecnico di Milano, Dipartimento di Elettronica e
  Informazione, 20133 Milano, Italy\\
  $^2$Research Laboratory of Electronics, Massachusetts Institute of
  Technology, Cambridge, Massachusetts 02139, USA\\
  $^3$Micro Photon Devices, via Stradivari 4, 39100 Bolzano, Italy}

\begin{abstract}
  The Rayleigh diffraction bound sets the minimum separation for two
  point objects to be distinguishable in a conventional imaging
  system. We demonstrate resolution enhancement beyond the Rayleigh
  bound using random scanning of a highly-focused beam and $N$-photon
  photodetection implemented with a single-photon avalanche detector
  array.  Experimental results show resolution improvement by a factor
  $\sim$$\sqrt{N}$ beyond the Rayleigh bound, in good agreement with
  theory.
\end{abstract}

\maketitle 
The response of a diffraction-limited imaging system to a point-like
source---its point spread function (PSF)---has extent inversely
proportional to the entrance pupil's area. The image is obtained by
convolving the system PSF with the light distribution of the object.
Therefore, details finer than the PSF's extent are lost under
conventional (entire-object) illumination. The Rayleigh diffraction
bound sets the minimum separation for two object points to be
distinguishable in the image.

Two classes of quantum strategies have been suggested to circumvent
this bound.  The first relies on techniques from quantum metrology
\cite{metrology}, in which the image information is encoded into
suitably tailored non-classical light
beams~\cite{review,shihimaging,lugiato}. For example, quantum
lithography \cite{litho} exploits the effective de Broglie
wave-function of $N$ photons in a delicately crafted state to obtain
an increase in resolution proportional to $N$ (see~\cite{mankei} for
recent developments). These methods are typically extremely sensitive
to photon loss and noise, because they rely on delicate quantum
effects (such as squeezing and entanglement).  Thus they are best
suited to short-distance applications, such as microscopy, where
losses can be controlled, as opposed to standoff sensing, such as
laser radar operation over km or longer path lengths, for which
substantial diffraction and atmospheric losses will be present.

The second class of quantum strategies for beating the Rayleigh
diffraction bound exploit postselection \cite{postsel} to extract the
high-resolution image associated with a non-classical component from
classical-state light containing information about the object to be
imaged
\cite{scullylith,agarwal,boyd,zhang,wang,korobkin,yablo,peer,jeff,qimaging}.
Because postselection involves discarding part of the measurement
data, these procedures inherently suffer detection inefficiency that
increases the time required to acquire an image.  However, their
spatial resolution can nonetheless exceed the Rayleigh diffraction
bound.  Furthermore, because they employ classical-state (laser)
light, these techniques degrade gracefully with increasing loss and
noise, making them suitable for standoff sensing.  In this Letter we
report the first experimental demonstration of one such technique,
viz., that of Giovannetti {\em et al.} \cite{qimaging}, in which the
object is illuminated by a focused light source and scanned. The
scanning pattern is irrelevant so long as the area of interest is
covered, i.e., a random scan pattern suffices.  The image is formed
using only those pixels that count exactly $N$ photons within a
measurement time $T$. The expected resolution improvement is
$\sqrt{N}$ over standard entire-object illumination, until the limit
set by the focused-beam illumination.  We begin with the theory for
this technique.

\section{Theory}
We are interested in an active imager, such as a laser radar,
comprised of a transmitter and a receiver in which we control the
object illumination and form the image with the receiver.  For such
systems, the spatial resolution is a function of two antenna patterns,
viz., the transmitter's illumination pattern on the object \em
and\/\rm\ the receiver antenna pattern, set by the diffraction limit
of its optics, projected onto the object.  When floodlight
illumination is employed, so that the entire object is bathed in
light, the resolution limit is set by the receiver's Rayleigh
diffraction bound.  If the transmitter and receiver are co-located,
they can share the same optics so that their antenna patterns have
identical Rayleigh bounds whose product gives the overall resolution
behavior.  Alternatively, if a small transmitter is located much
closer to the object than is the receiver, it is possible to project
very small spots onto the object to be imaged.  If this is done in a
precision scan, so that the receiver knows exactly where the
transmitter is pointing at any instant in time, a simple energy
measurement at the receiver will realize resolution limited by the
transmitter's antenna pattern, regardless of the receiver's own
Rayleigh bound.  However, creating that precision scan, and relaying
the scan positions to the receiver, could easily be a major challenge,
especially if the transmitter is mounted on a small unmanned air
vehicle.  The postselection technique of Giovannetti \em et al.\/\rm\
circumvents that problem by allowing the scan pattern to be arbitrary,
even random, so long as the object's region of interest is covered.

Suppose that a focused transmitter emits a $+z$-going,
quasimonochromatic, paraxial, linearly-polarized laser pulse with
scalar complex envelope $E_T(\rhovec,t;\thetavec) = \sqrt{4N_T/\pi
  D_T^2}\,s(t)e^{-ik|\rhovec|^2/2L_T + ik\thetavec\cdot\rhovec}$ for
$|\rhovec| \le D_T/2$, where $\rhovec = (x,y)$ is the transverse
coordinate vector, $\thetavec = (\theta_x, \theta_y)$ is the
transmitter aim angle, and $k$ is the wave number at the center
wavelength $\lambda$.  We will normalize $E_T$ to have units
$\sqrt{\mbox{photons/m$^2$s}}$, and take the pulse shape $s(t)$ to
satisfy $\int\!dt\,|s(t)|^2 = 1$, so that $N_T$ is the average
transmitted photon number \cite{footnote1}.  This pulse
transilluminates an object $L_T$-m away from the transmitter
\cite{footnote2}.  The light that passes through the object is then
collected by a diffraction-limited circular lens of diameter $D_R$
located $L_R$-m in front of the object.  The focal length of this lens
is such that it casts an image of the object at a distance $L_I > L_R$
beyond the lens.  Neglecting the propagation delay and correcting for
image inversion, the photon-flux density in the image plane is then
\begin{eqnarray}
\lefteqn{|E_{\rm IM}(\rhovec_{\rm IM},t;\thetavec)|^2 = N_T|s(t)|^2
  \times} \nonumber \\ &&\left|\int\,d\rhovec\,
{\cal{O}}(\rhovec)\sqrt{\frac{\pi D_T^2}{4(\lambda L_T)^2}}
\frac{\pi D_R^2}{4\lambda^2 L_RL_I}\,e^{ik|\rhovec|^2(L_T^{-1} +
  L_R^{-1})/2}\right. \times\nonumber \\
&&\hspace*{-.1in}\left.\frac{J_1(\pi D_T|\rhovec - \thetavec
    L_T|/\lambda L_T)}{\pi D_T |\rhovec - \thetavec L_T|/2\lambda
    L_T}
\frac{J_1(\pi D_R|\rhovec - \rhovec_{\rm IM}/m|/\lambda L_R)}{\pi
  D_R|\rhovec - \rhovec_{\rm IM}/m|/2\lambda  L_R}\right|^2, \nonumber
\end{eqnarray}
where ${\cal{O}}(\rhovec)$ is the object's field-transmission
function, $m \equiv L_I/L_R$ is the image magnification
\cite{footnote3}, and we see the transmitter and receiver's
circular-pupil antenna patterns.

To exhibit the sub-Rayleigh resolution capability of the scheme from
\cite{qimaging} we shall assume that: (1) the transmitter's antenna
pattern fully resolves all significant features in
${\cal{O}}(\rhovec)$ \cite{footnote4}; (2) the image-plane photon
counting array has pixels of area $A_p$ sufficiently small that
diffraction, rather than pixel size, limits image resolution; and (3)
the photon counting array outputs pixel counts taken over the full
$T$-s-long extent of $s(t)$.  For a given illumination angle
$\thetavec$, the pixel counts are then statistically independent,
Poisson random variables with mean value, for the pixel centered at
$\rhovec_{\rm IM}$, given by
\begin{eqnarray}
  \bar{N}_{\thetavec}(\rhovec_{\rm IM}) &=& \eta
  N_T|{\cal{O}}(\thetavec L_T)|^2\frac{\pi
    D_R^4L_T^2A_p}{4D_T^2\lambda^2 L_R^2L_I^2} \nonumber \\ 
  &\times& \left(\frac{J_1(\pi D_R|\thetavec L_T - \rhovec_{\rm
        IM}/m|/\lambda L_R)}{\pi D_R|\thetavec L_T - \rhovec_{\rm
        IM}/m|/2\lambda  L_R}\right)^2,  \nonumber
\end{eqnarray}
where $\eta$ is the detector's quantum efficiency.  If photon counts
are collected from the pixel at $\rhovec_{\rm IM}$ while $\thetavec$
is randomly scanned over the object region, the unconditional
probability of getting $N$ counts from that pixel is
\begin{equation}
  P_N(\rhovec_{\rm IM}) = 
\int\!d\thetavec\,p(\thetavec) \frac{\bar{N}_{\thetavec}(\rhovec_{\rm IM})^N 
    e^{-\bar{N}_{\thetavec}(\rhovec_{\rm IM})}}{N!}, \nonumber
\end{equation}
where $p(\thetavec)$ is the scan pattern's probability density
function.  Postselecting those pixels for which $N$ counts have been
registered, we get an image ${\cal{I}}_N(\rhovec_{\rm IM}) \propto
P_N(\rhovec_{\rm IM})$.  For $N > \max\bar{N}_{\thetavec}(\rhovec_{\rm
  IM})$, the Poisson distribution is monotonically decreasing with
increasing $N$, whence
\begin{eqnarray}
  {\cal{I}}_{N}(\rhovec_{\rm IM}) &\sim&
  \int\!d\thetavec\,p(\thetavec)|{\cal{O}}(\thetavec L_T)|^{2N}
  \nonumber \\
  &\times& \left(\frac{J_1(\pi D_R|\thetavec L_T - \rhovec_{\rm
        IM}/m|/\lambda L_R)}{\pi D_R|\thetavec L_T - \rhovec_{\rm
        IM}/m|/2\lambda L_R}\right)^{2N}, \nonumber
\end{eqnarray}
where we have suppressed multiplicative constants and ignored the
exponential term as it is independent of $N$.  Here we see that the
postselected image contains $|{\cal{O}}|^{2N}$ convolved with the
$N$th power of the receiver's Airy disk pattern, i.e., a point-spread
function that is $\sim$$\sqrt{N}$ narrower than the Airy disk itself
when compared on a main-lobe area basis \cite{qimaging}.  Note that
violating the preceding monotonicity condition can lead to the
``donut-hole'' effect exhibited in the experiments described below.

\section{Experiment}
We demonstrate the concept of sub-Rayleigh imaging with the setup
shown in Fig.~\ref{f:setup}(a). The object to be imaged in
transmission was part of a U.S.\ Air Force (USAF) resolution target
consisting of alternate opaque and clear stripes of width 125\,$\mu$m
(4 line pairs/mm), as indicated by the arrow in Fig.~\ref{f:image}(a).
A 532-nm laser was mounted on an $XY$ translation stage that provided
scan coverage over the entire object with a 20-$\mu$m-radius focused
spot. We imaged the object through a $f$ = 25-cm diffraction-limited
lens set in a 2-mm-diameter aperture.  The optics provided 5.3$\times$
image magnification, yielding a 660-$\mu$m-wide strip at the image
plane. Under conventional (entire-object) illumination, shown in the
setup of Fig.~\ref{f:setup}(b), the Rayleigh diffraction bound for the
imaging system at the image plane was 1.86\,mm, which is
$\sim$2.8$\times$ larger than the stripe width.
Figure~\ref{f:image}(b) shows the conventional-illumination image that
was obtained using standard photodetection (with all events counted):
the stripes are unresolved, as expected. Note that with full
illumination of the object, we were not able to go beyond the Rayleigh
bound even with the $N$-photon detection scheme, indicating that
focused illumination is a necessary requirement.

\begin{figure}[h!]
\centerline{\epsfxsize=.8\hsize\epsffile{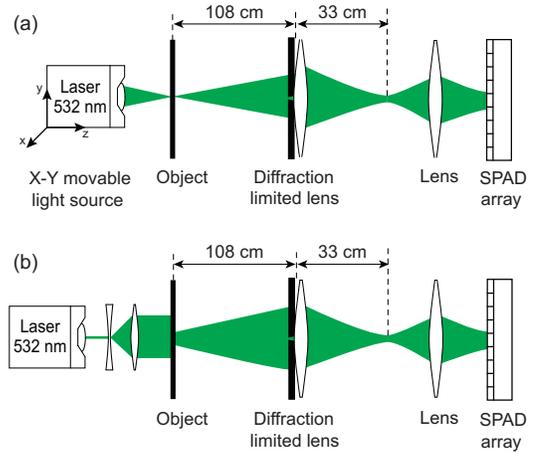}}
\caption{(color online). Setup schematics for (a) sub-Rayleigh imaging
  with focused illumination, and (b) conventional coherent imaging
  with full illumination.
  \labell{f:setup}}
\end{figure}

The detector was a compact $32\times 32$ Si single-photon avalanche
diode (SPAD) array fabricated with a complementary metal oxide
semiconductor (CMOS) process \cite{spad}.  The Si SPAD was a $p$-$n$
junction reverse-biased above its breakdown voltage and operated in
the Geiger mode with a detection efficiency of $\sim$30\% at 532\,nm.
Each pixel of the CMOS SPAD array consisted of one SPAD with its
front-end active quenching and resetting electronics and a digital
counting circuitry for in-pixel pre-processing.  The pixel pitch for
the array was 100\,$\mu$m, and the SPAD had a fill factor of only
3.1\% at each pixel due to the presence of on-chip electronics.  Owing
to the large separation between SPADs we did not observe any cross
talk. The average pixel dead time (including all the electronic
circuitry) was 300\,ns. Each SPAD delivered a digital output pulse for
every single-photon detection event with no readout noise. The
in-pixel counting circuitry would compute the number of single-photon
events within its user-selectable integration time of 1\,$\mu$s or
more and store the tally in an in-pixel memory cell. By measuring
incident photons over a long integration time, $N$-photon sensitivity
in the time domain can be achieved at the single-pixel level. The
array readout was performed through an 8-bit data bus without
interrupting the next 1024-pixel frame of photon-counting integration,
and the maximum frame rate was $10^5$/s. A typical integration time of
each frame was tens of $\mu$s.

\begin{figure}[h!]
\centerline{\epsfxsize=1.\hsize\epsffile{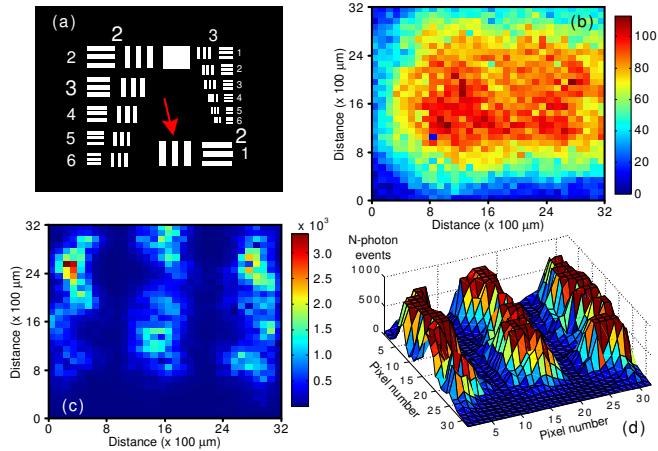}}
\caption{(color online).  (a) Object indicated by arrow: target \#\,1,
  group 2 of a USAF resolution target, composed of opaque and clear
  stripes 125\,$\mu$m wide. (b) Blurred image obtained conventionally
  using full illumination and taken in a single 50-$\mu$s frame;
  Rayleigh diffraction bound is 1.86\,mm. (c) Sub-Rayleigh image using
  focused illumination and $N$ = 23. Details are obscured by a few
  pixels with very high event counts. (d) 3-D intensity profile of (c)
  with the stripes clearly revealed by clipping pixels with very high
  event counts at the limit of 800.
  \labell{f:image}}
\end{figure}

To implement sub-Rayleigh imaging in Fig.~\ref{f:setup}(a) we manually
scanned the focused beam in a random pattern, making sure that there
was coverage for the entire area of interest. At each scan location,
we recorded over 8000 measurement frames for image averaging that took
less than 1 second to accomplish. The incident power was adjusted to
have an average peak photocount $\bar{N}_{\rm peak}$ =14 per
integration time (for one pixel). For each measurement frame, each
pixel with exactly $N$ photocounts (after dark-count subtraction,
measured separately) was tagged as having an $N$-photon event.  All
other pixels were then tagged for zero $N$-photon events. The
measurement process was then repeated at a different scan location
until the object of interest was fully scanned.
Figure~\ref{f:image}(c) shows the resultant image for $N$ = 23,
revealing the three stripes that were lost under conventional
illumination in Fig.~\ref{f:image}(b). The color scale of
Fig.~\ref{f:image}(c) has a large range to accommodate several pixels
with very high event counts and therefore image details (with lower
event frequencies) are obscured.  Figure~\ref{f:image}(d) shows the
3-D intensity profile of the same $N=23$ image of
Fig.~\ref{f:image}(c), except that we cap the event occurrence at 800
for pixels with over 800 events in order to make the lower-count
pixels more visible, thus revealing the three stripes very clearly.
According to theory, the expected enhancement of $\sqrt{N}$ should
yield a sub-Rayleigh resolution of $\sim$1.86/$\sqrt{23}$ = 0.4\,mm is
qualitatively borne out by our results.  Note that the sub-Rayleigh
resolution still exceeds the 106-$\mu$m-limit set (after
magnification) by the focused illumination at the object, as expected.

We chose $N=23$ to be substantially larger than $\bar{N}_{\rm peak}$
=14 to avoid the ``donut-hole" problem.  To illustrate this issue,
Fig.~\ref{f:pointsource} shows images of a point source obtained under
various measurement conditions. The aperture diameter of the imaging
optics in Fig.~\ref{f:setup}(a) was set to 3\,mm with the same overall
image magnification of 5.3 so that the Rayleigh bound at the image
plane (SPAD array) was 1.2\,mm. We removed the USAF resolution target
so that the 20-$\mu$m-radius focused spot at the object plane served
as the point source, and no scanning was necessary for these images.
We took $\sim$32,000 measurement frames, recorded the photocounts at
each pixel for each frame, binned them accordingly after subtracting
dark counts, and processed the data.  Figure~\ref{f:pointsource}(a)
shows the cross section of the Rayleigh-bound image of the point
source through the 3-mm-diameter aperture obtained by including all
photocounts to yield an intensity profile averaged over the
$\sim$32000 frames. We measured $\bar{N}_{\rm peak}\approx 15$ and, as
an indicator for the image size, we obtained the full-width at
half-maximum (FWHM) of $\sim$1\,mm.

\begin{figure}[h!]
\centerline{\epsfxsize=.9\hsize\epsffile{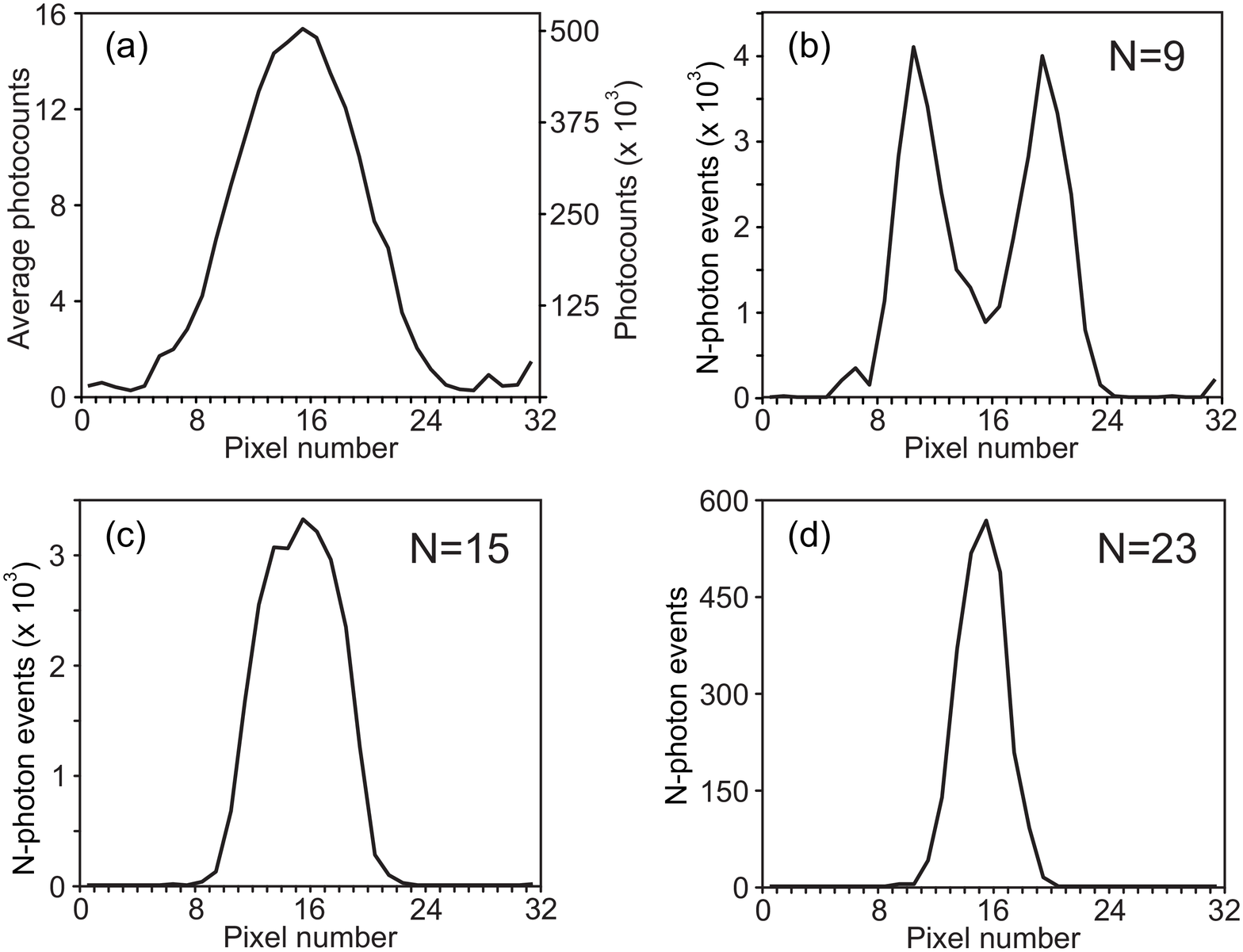}}
\caption{Image cross section of a point source with a modified setup
  of Fig.~\ref{f:setup}(a) (3-mm aperture) obtained by collecting: (a)
  all photocounts as in conventional imaging with $\bar{N}_{\rm
    peak}\approx 15$; (b) exactly $N$ =9  photocounts and showing a
  hole where a peak should be, (c) exactly $N$ =15 photocounts with a
  peak slightly sharper than in (a); and (d) exactly $N$ =23
  photocounts with a sub-Rayleigh peak that is sharper than (c). 
  \labell{f:pointsource}}
\end{figure}

Figures~\ref{f:pointsource}(b)--(d) are cross-sectional profiles
obtained by selecting exactly $N$-photocounts for $N=$ 9, 15, and 23,
that are respectively smaller than, equal to, and greater than
$\bar{N}_{\rm peak}$. For $N<\bar{N}_{\rm peak}$ in (b), the center
portion of the point-source image usually received more than the
threshold level $N$ and therefore had few exactly $N$-photon events.
On the other hand, the photocounts away from the center decrease from
$\bar{N}_{\rm peak}$ until the photocount average matches the
threshold $N$ where it shows a peak, and hence the image has a
``donut-hole" shape \cite{Migdall}.  For $N\approx \bar{N}_{\rm peak}$
in (c), the profile is single-peaked and looks slightly narrower and
steeper than the Rayleigh-bound image in Fig.~\ref{f:pointsource}(a)
which is much larger than the magnified size of the focused beam of
125 $\mu$m (FWHM) at the image plane. When $N\gg\bar{N}_{\rm peak}$,
as required for sub-Rayleigh imaging, we observe in (d) that the
profile for $N$ = 23 shows a much narrower peak with a FWHM width of
$\sim$0.4\,mm that is smaller than the 1-mm width (FWHM) of the
Rayleigh-bound point source image in Fig.~\ref{f:pointsource}(a).  We
also note that the $N$ = 23 event frequency is much lower than for $N$
less than or equal to $\bar{N}_{\rm peak}$, because such a large $N$
did not happen very often. Our sub-Rayleigh imaging has
characteristics that are similar to $N$-photon interferometry, in
which sub-wavelength ($\lambda/N$) interference patterns were obtained
with coherent-state input and postselective $N$-photon detection
\cite{Bouwmeester}.

Our point-source images suggest the following physical origin for the
formation of the Rayleigh diffraction bound and the basis for our
sub-Rayleigh imaging technique.  The donut holes that form for
$N<\bar{N}_{\rm peak}$ are responsible for the diffractive spread of
the imaging system.  By removing the low-$N$ photocount events and
those comparable to $\bar{N}_{\rm peak}$, one captures the much
narrower profile for $N\gg\bar{N}_{\rm peak}$ at the expense of longer
acquisition times due to less frequent occurrences for high values of
$N$.  We should also achieve similar results if we relax the
measurement requirement from exactly $N$ for $N\gg\bar{N}_{\rm peak}$
to a sum of all $N\gg\bar{N}_{\rm peak}$, whose resolution should be
dominated by its lowest $N$ term due to its larger sub-Rayleigh width
($\propto 1/\sqrt{N}$) and higher rate of occurrence.

In conclusion, we have demonstrated sub-Rayleigh imaging resolution
using a classical light source, tight focusing on the object, and
$N$-photon photodetection.  The sub-Rayleigh technique removes the
low-$N$ image components that contribute to the diffractive spread of
the imaging system. The measured resolution enhancement of
$\sim$$\sqrt{N}$ is in good agreement with theory. The sub-Rayleigh
technique may find applications in which active illumination can be
implemented to obtain a higher resolution image at the expense of
longer acquisition times.

We thank J.~Le Gou\"{e}t for help with the experiment. F.~G.
acknowledges the Roberto Rocca Doctoral Fellowship. This work was
supported in part by the W.\ M.~Keck Foundation Center for Extreme
Quantum Information Theory, the U.S.\ Army Research Office under a
Multidisciplinary University Research Initiative grant, and the DARPA
Quantum Sensors Program.


\begin{references}
\bibitem{footnote2} Standoff sensing is ordinarily done in reflection,
  but we assume transillumination here to match our experiment.
\bibitem{footnote3}Laser radar operation would employ $L_I < L_R$,
  hence image minification. We have specified $L_I > L_R$ to match our
  experiment.
\bibitem{footnote4}If $d$ is the minimum size of any significant
  ${\cal{O}}(\rhovec)$ feature, then we require $dD_T/\lambda L_T \gg
  1$.
\end{references}
\end{document}